\documentclass[conference]{IEEEtran}
\IEEEoverridecommandlockouts

\usepackage{cite}

\usepackage{tikz}
\usepackage{amsmath}

\usepackage{filecontents}

\usepackage{amsmath}           
\usepackage{amssymb}
\usepackage{amsfonts}
\usepackage{bm} 

\usepackage{algorithm}
\usepackage{algpseudocode}

\usepackage{graphicx}          
\usepackage[center]{caption}
\usepackage{subcaption}
\usepackage{float}
\usepackage{booktabs}
\usepackage{multirow}
\usepackage{multicol}

\usepackage{ifsym}
\usepackage{marvosym}

\usepackage{framed}
\usepackage{textcomp}
\usepackage{xcolor}
\usepackage{url}
\usepackage{setspace}
\usepackage{comment}
\usepackage{hyperref}

\def\BibTeX{{\rm B\kern-.05em{\sc i\kern-.025em b}\kern-.08em
    T\kern-.1667em\lower.7ex\hbox{E}\kern-.125emX}}

\usepackage[normalem]{ulem}

\begin{document}

\title{Psyzkaller: Learning from Historical and On-the-Fly Execution Data for Smarter OS Kernel Fuzzing}

\author{
    \IEEEauthorblockN{Boyu Liu\IEEEauthorrefmark{1}, Yang Zhang\IEEEauthorrefmark{2}, Liang Cheng\IEEEauthorrefmark{3}, Yi Zhang\IEEEauthorrefmark{4}, Junjie Fan\IEEEauthorrefmark{5}, Xiaoshan Sun\IEEEauthorrefmark{6}, Yu Fu\IEEEauthorrefmark{7}, Zhen Li\IEEEauthorrefmark{8}, Dengguo Feng\IEEEauthorrefmark{9}}
    
    \IEEEauthorblockA{\IEEEauthorrefmark{1}\IEEEauthorrefmark{5}University of Chinese Academy of Sciences\\Beijing, China}
    
    \IEEEauthorblockA{\IEEEauthorrefmark{1}\IEEEauthorrefmark{2}\IEEEauthorrefmark{3}\IEEEauthorrefmark{5}\IEEEauthorrefmark{6}\IEEEauthorrefmark{9}TCA Lab, Institute of Software, Chinese Academy of Sciences\\Beijing, China\\Email: \{liuboyu2024,  zhangyang, chengliang, junjie2020, sunxs, fengdg\}@iscas.ac.cn}
    
    \IEEEauthorblockA{\IEEEauthorrefmark{2}Zhongguancun Lab\\Beijing, China}
    
    \IEEEauthorblockA{\IEEEauthorrefmark{4}\IEEEauthorrefmark{7}Cryptography and Cyberspace Security (Whampoa) Institute\\Guangzhou, China\\Email: \{zhangyi, fuyu\}@ioccs.cn}

    \IEEEauthorblockA{\IEEEauthorrefmark{8}School of Cyber Science and Technology, Shandong University\\Jinan, China\\Email: lizhen202221104@mail.sdu.edu.cn}

    \thanks{\IEEEauthorrefmark{2}\IEEEauthorrefmark{3}\IEEEauthorrefmark{4}Corresponding authors}
}

\maketitle

\begin{abstract}



OS Kernel fuzzers such as Syzkaller often struggle to generate syscall sequences that respect intrinsic Syscall Dependency Relations (SDRs), resulting in seeds that either violate kernel constraints or fail to reach deep execution paths.

We propose leveraging an N-gram model to learn SDRs from both kernel execution history and ongoing fuzzing results. This enables the fuzzer to capture dependencies in similar kernel versions while adapting to target-specific behaviors, thereby improving the validity of generated seeds.  Additionally, we introduce a bidirectional Random Walk strategy to enhance the diversity of generated seeds.

We implement this approach in a prototype, Psyzkaller, on top of Syzkaller. Experiments show that,  trained with the large-scale DongTing dataset and continuously updated with ongoing fuzzing results, Psyzkaller improves Syzkaller’s code coverage by 4.6\%–7.0\%, triggers 110.4\%–187.2\% more crashes, and discovers eight previously unknown kernel vulnerabilities.  Furthermore, Psyzkaller outperforms state-of-the-art fuzzers such as ACTOR and SyzDescribe in both coverage and crashes.

\end{abstract}


\section{Introduction}

Fuzzing has emerged as one of the most effective automated testing techniques for OS kernels. Despite its widespread adoption, state-of-the-art (SOTA) kernel fuzzers typically faces a challenge that significantly limits their effectiveness: they often generated syscall sequences (SCSs) that violate the intricate  \emph{syscall dependency relations} (SDRs),  i.e., the semantic constraints governing valid interactions among syscalls. For instance, a generated SCS may invoke \texttt{write()} to write a file before calling \texttt{open()} to obtain the corresponding file descriptor. Such violations cause the kernel to reject the SCSs or force the fuzzer to prematurely terminate its exploration of the kernel’s code space, resulting in great loss in efficiency. 

Previous research has attempted to infer SDRs either through static analysis of kernel code~\cite{corina_difuze:_2017, han_imf:_2017} or by applying machine learning (ML) techniques to fuzzing execution data~\cite{chen_syzgen_2021, kim_hfl_2020, bulekov_no_2023}, and then enforcing the inferred SDRs when generating SCSs. While promising, these approaches remain computationally expensive, making them difficult to integrate into the high-throughput workflow of kernel fuzzing. Moreover, the available fuzzing execution data is often insufficient for ML-based methods to fully train their models and offer useful guidance on seed generation.

To address this challenge, we propose \emph{learning} SDRs from both historical kernel executions and ongoing execution traces generated during fuzzing, as they provide concrete evidence of how syscalls interact in the target kernel and similar kernel versions. Learning from historical data offers broad coverage of SDRs but may face compatibility issues when applied to the target kernel. In contrast, on-the-fly learning captures dependencies specific to the target kernel but is limited to the portions of the code space already explored, especially during the early stages of fuzzing. We argue that combining these two sources achieves the best of both worlds: historical data provides comprehensive prior knowledge, while online learning enables adaptability and continuous refinement throughout fuzzing.

Realizing this vision requires addressing three key requirements: (1) the learned SDRs must be interpretable and inspectable; (2) the learning process must be computationally efficient to keep pace with the rapid SCS generation of fuzzing; and (3) the learned model must integrate smoothly with existing fuzzing workflows.

To this end, we design and implement \textbf{Psyzkaller}, an extension of Syzkaller that learns SDRs using an Bigram language model~\cite{Habash04}. This offers three distinct advantages. (1) \emph{Interpretability:} SDRs are encoded as transition probabilities between syscalls in the N-gram model, making them easy to understand and inspect. (2) \emph{Efficiency:} learning SDRs is executed as updating the parameters of a Bigram model—essentially a matrix—that is computationally lightweight compared to other neural language models. (3) \emph{Seamless integration:} the structure of the Bigram model directly matches Syzkaller’s Choice Table, allowing Psyzkaller to guide seed generation without modifying Syzkaller’s core mechanisms. In addition, we introduce an optimization called RandomWalk that enables Syzkaller to generate SCSs bi-directionally based on the learned SDRs to improve the diversity of generated seeds. 

We pre-train Psyzkaller using the DongTing dataset~\cite{dongting}, one of the most comprehensive Linux kernel execution trace datasets publicly available, and compare its fuzzing performance against several openly available SOTA kernel fuzzers including Syzkaller,  ACTOR\cite{actor}, and SyzDescribe\cite{syzdescribe}, on a group of currently widely-used Linux kernel versions (5.19, 6.1, 6.2 and 6.12).  Our experiments show that Psyzkaller surpassed Syzkaller by 4.1\%--7.0\% on branch coverage and by 57.9\%--187.2\% on the number of crashes triggered, depending on kernel version. It also outperformed ACTOR and SyzDescribe. Additionally, Psyzkaller detected eight (8) previously unknown vulnerabilities, as compared to only one (1) by Syzkaller.

In summary, this paper makes the following contributions:
\begin{itemize}
    \item We propose a novel strategy of learning SDRs from both historical and ongoing kernel execution traces generated during fuzzing to guide seed generation in kernel fuzzing. 
    \item We develop Psyzkaller, which integrates Bigram-based SDR learning, as well as the RandomWalk strategy, into Syzkaller’s workflow.  
    \item We evaluated the effectiveness and efficiency of Psyzkaller against SOTA kernel fuzzers, as well as its best learning strategy on historical data, with experiments across four recent, representative Linux kernel versions. 
\end{itemize}

The rest of this paper is organized as follows: Section~\ref{sec:Background} provides background information pertain to our approach;  Section~\ref{sec:approach} describes our approach in details, followed by implementation details in Section~\ref{sec:Implementation}; Section~\ref{sec:Evaluation} presents the evaluation of Psyzkaller; Section~\ref{sec:Discussion} discusses a few considerations in the evaluation;  Section~\ref{sec:related-work} compares our approach to related research; lastly, Section~\ref{sec:Conclusion} concludes the paper. 
\section{Background}\label{sec:Background}
\subsection{N-gram}
N-gram~\cite{Bengio03} is a type of statistical model widely used in natural language processing (NLP), due to its strong interpretability, high computational efficiency, and applicability to small datasets. It works on the assumption that the probability of word $w_i$ appearing in a sentence only depends on the proceeding $N-1$ words ($N\in [2,3,4..]$). That is, given a prefix $w_1, w_2, ..., w_{i-1}$, the probability that the next word is $w_i$ is: $ p(w_i \mid w_{1}, \ldots, w_{i-1}) = p(w_i \mid w_{i-N+1}, \ldots, w_{i-1}) $.

An N-gram model learns the probability distribution of words conditioned on their preceding context, enabling a range of NLP tasks such as text generation, classification, and spelling correction. Among the N-gram variants, the Bigram model ($N = 2$) is particularly popular due to its simplicity and low computational cost. In a Bigram model, the probability of observing a word $w_i$ depends solely on its immediate predecessor, formalized as $p(w_1,...,w_K) = p(w_1) \cdot \prod_{i=2}^K{p(w_i|w_{i-1})}$.


In this work, we adopt the Bigram model to learn SDRs from historical and on-the-fly kernel execution data, balancing computational cost with the amount of data required for training.  Although higher-order N-gram models (e.g., Trigram with $N=3$) may be able to capture more complex SDRs—particularly those spanning longer syscall sequences—they demand substantially greater computational effort and significantly larger datasets to produce reliable statistics.

\subsection{The DongTing Dataset}\label{ssec:dt}
In this work, we use the DongTing dataset~\cite{dongting} as the source of historical Linux kernel execution data to train Bigram because of its comprehensiveness and public accessibility. Compared to other available datasets, it is more comprehensive in terms of coverage on kernel versions and kernel executions: it comprises 85 GB of data, including 18,966 syscall sequences collected from more than 200 Linux kernel versions (4.13–5.17).

Dongting is composed of 6,850 \emph{normal} traces and  12,116 \emph{abnormal} traces, where the former are Linux kernel executions that successfully completed and the latter are those leading to crash or hangs. Specifically, normal traces in the Dongting dataset were collected from executing four widely-used Linux regression test suites (namely LTP~\cite{linux_ts}, Kselftests~\cite{kern_ts}, Glibc~\cite{glibc_ts}, and the Open POSIX Test Suite~\cite{posix_ts}), while abnormal traces were extracted from executing proof-of-concept (PoC) programs for crashes reported on the Syzbot dashboard~\footnote{Syzbot is a Syzkaller community platform for contributors to report Linux kernel crashes.}. 

\subsection{Syzkaller}\label{ssec:syz}
Syzkaller~\cite{syz}, an open-source fuzzer developed and maintained by Google, has become the de facto tool for fuzzing kernels such as Linux and FreeBSD. As illustrated in the yellow entities of Figure~\ref{fig:archi}, it consists of two main components: (i) \texttt{syz-manager}, which orchestrates the fuzzing session by generating and mutating SCS seeds, maintaining the seed queue, launching and monitoring virtual machine (VM) instances, and collecting crash reports; and (ii) \texttt{syz-executor}, which runs within each VM to execute test cases on the target kernel and collect feedback such as code coverage and error states.

Syzkaller generates SCSs using a Choice Table, a two-dimensional matrix encoding the probability that one syscall follows another. Seed generation starts with a randomly selected syscall, after which each subsequent syscall is chosen according to the cumulative probabilities in the Choice Table, increasing the likelihood that the sequence passes kernel validation and explores unseen code paths.

The Choice Table is constructed from two complementary sources. First, analysis of syscall templates—written by human experts in the \texttt{syzlang} language to specify syscall interfaces—and constants extracted from kernel source code identifies candidate syscall relationships (e.g., via parameter types or return values). Second, dynamic feedback from actual executions captures realistic syscall-ordering patterns. Syzkaller combines these static and dynamic signals to balance syntactic interface knowledge with empirical execution behavior.



\section{Approach}\label{sec:approach}
The core idea of learning more accurate and comprehensive SDRs from both historical kernel execution data and ongoing traces generated during fuzzing is based on the assumption that syscalls with dependency relations tend to appear consecutively in real executions. This strategy balances the broad coverage of SDRs present across OS kernel versions with the dependencies unique to the target kernel. Such balanced SDRs help the fuzzer explore the kernel’s code space both broadly and deeply. Since our approach focuses on enforcing the order of syscalls in generated seeds to comply with the learned SDRs, it leverages Syzkaller’s existing mechanisms to assign parameter values. Investigating whether syscall-parameter dependencies inferred from execution data could further improve or replace Syzkaller’s current parameter-generation strategy is left to future work.

\subsection{Learning SDRs with N-gram}
Our approach uses an N-gram model, rather than deep neural networks (DNNs) or other AI-based methods, for SDR learning for three reasons. First, DNN-based models such as Word2Vec~\cite{mikolov2013efficient}, recurrent neural networks (RNNs)~\cite{medsker2001recurrent}, and Transformers~\cite{NIPS2017_3f5ee243} typically require extremely large training corpora—on the order of millions of tokens—to achieve satisfactory performance. Such datasets however are not available in the kernel fuzzing domain. In contrast, N-gram models can achieve reasonable accuracy with substantially smaller datasets, making them well-suited for learning SDRs from modest corpora, which usually consist of thousands to tens of thousands of kernel execution traces~\cite{pailoor_moonshine:_2018}, while also allowing continuous learning from on-the-fly fuzzing data.


Second, N-gram models are computationally efficient, require no complex hyperparameter tuning, and can be easily updated with new fuzzing results through simple matrix adjustments. In contrast, DNN-based models have millions of parameters—for example, a 5,000-token Word2Vec model has roughly 5000×128×2=1.28 million parameters. Scaling such models to cover Linux syscalls and their SDRs (roughly 500 Linux syscalls and around 100,000 SDRs) is computationally expensive and prone to overfitting.


Finally, N-gram models provide better interpretability: SDRs are represented directly as conditional probabilities among syscalls within the Bigram matrix, enabling straightforward inspection and analysis. This structure also makes it convenient to cross-check inferred SDRs against syscall documentation or kernel source code, thereby improving their accuracy and correctness.

%
When being used to learn SDRs,  Bigram essentially formulates a matrix, referred to as the Relation Probability Matrix or $RPM$, in which $RPM[i][j]$ records $p(w_j|w_i)$—the probability of $w_j$ being immediately invoked after $w_i$ is invoked. According to the Bayes’ theorem and Bigram equation, $p(w_j |w_i)$ can be estimated as $ {\sf count}(w_i w_j) / {\sf count}(w_i)$, where ${\sf count}($g$)$ returns the number of appearance of syscall or syscall sequence $g$ in the kernel execution corpus $C$. 


Specifically, given the input Corpus $C$, Algorithm~\ref{alg:learningSDR} computes the trained $RPM$. After initializing the $RPM$ (Lines~\ref{alg:sdr:start_init}-\ref{alg:sdr:end_init}),  the algorithm traverses $C$ to count the occurrences of individual syscalls and consecutive syscall pairs (Lines~\ref{alg:sdr:start_trav}-\ref{alg:sdr:end_trav}), and then calculates $RPM$ based on these counts (Lines~\ref{alg:sdr:start_calc}-\ref{alg:sdr:end_calc}). In Algorithm~\ref{alg:learningSDR},  `$|sc|$' at Line~\ref{alg:sdr:len_calc} returns the length of SCS $sc$, and $IndexOfCall()$ returns the index of a syscall.  Notably,  $RPM$ reuses the syscall indices from Syzkaller’s Choice Table, facilitating seamless integration of the two during seed generation. Consequently, the size of  $RPM$ matches that of Syzkaller’s Choice Table.

\begin{algorithm}
\caption{LearnSDR}
\label{alg:learningSDR}
\begin{algorithmic}[1]
\Require 
    $C$: Corpus of syscall sequences;
    $CallNum$: Total number of syscalls in the kernel.
\Ensure 
    $RPM$: Bigram model learned from $C$.

\For{$i = 0$ to $CallNum - 1$} \label{alg:sdr:start_init}
    \For{$j = 0$ to $CallNum - 1$}
        \State $C[i][j] \gets 0$
    \EndFor
    \State $Count[i] \gets 0$
\EndFor\label{alg:sdr:end_init}

\ForAll{$sc \in C$}\label{alg:sdr:start_trav}
    \For{$i = 0$ to $|sc| - 2$} \label{alg:sdr:len_calc}
        \State $x \gets {\sf IndexOfCall}(sc[i])$
        \State $y \gets {\sf IndexOfCall}(sc[i+1])$
        \State $RPM[x][y] \gets RPM[x][y] + 1$
        \State $Count[x] \gets Count[x] + 1$
    \EndFor
\EndFor\label{alg:sdr:end_trav}

\For{$i = 0$ to $CallNum - 1$}\label{alg:sdr:start_calc}
    \For{$j = 0$ to $CallNum - 1$}
        \If{$Count[i] \neq 0$}
            \State $RPM[i][j] \gets RPM[i][j] / Count[i]$
        \EndIf
    \EndFor
\EndFor\label{alg:sdr:end_calc}

\end{algorithmic}
\end{algorithm}

\subsection{Selecting Historical Dataset for Learning SDRs}
The quality of historical kernel execution datasets directly affects the coverage and fidelity of learned SDRs, and consequently their impact on improving fuzzing effectiveness.  Nonetheless, our experiments show that even small historical datasets—or learning SDRs solely from ongoing fuzzing—can already enhance Syzkaller’s performance (see Section~\ref{sec:Evaluation}).  

An ideal historical corpus should provide broad coverage of latent SDRs in the Linux kernel, typically measured by both the number of executions and the range of kernel versions, and include both normal and abnormal traces. Normal traces support wide exploration of the kernel’s execution space, while abnormal traces often reveal critical SDRs linked to crashes and vulnerabilities.

We conducted a comprehensive survey of syscall sequence datasets of realistic Linux kernel executions and evaluated their suitability for training the Bigram model based on three criteria: size, public availability, and the presence of both normal and abnormal traces. Many historical datasets reported in the literature were originally created for tasks such as network intrusion detection or host anomaly detection and are therefore unsuitable for learning SDRs.  Table~\ref{tab:dataset} summarizes four evaluated datasets. Firefox-DS~\cite{firefoxds} and ADFA-LD~\cite{ADFA-LD} are small (fewer than 6,000 traces), outdated (both released in 2013), and not publicly available. PLAID~\cite{PLAID} primarily contains traces from network services, limiting coverage of syscalls related to other kernel components such as the file system. Based on this evaluation, the DongTing dataset was selected as it outperforms the others across all three criteria.

\begin{table}[htbp]
\centering
\small{
\caption{Historical Linux Kernel Execution Datasets}
\label{tab:dataset}
\begin{tabular}{llllll}
\toprule
\textbf{Dataset} & \textbf{Size} & \multicolumn{2}{c}{\bf Seq. Count} &\textbf{Avail} & \textbf{Released}\\
\cmidrule{3-4}
  & {\bf (GB)} & \textbf{Norm.} & \textbf{Abn.} &\textbf{-ability} & \textbf{Date}\\
\midrule
FirefoxDS  &0.27 & 700 & 19 & $\times$& Nov, 2013\\
ADFA-LD  & 0.016 & 5,206 & 746 & $\times$& Jan, 2013\\
PLAID  & 0.78 & 39,672 & 1145 & \checkmark& Oct, 2020\\
DongTing  & 85 & 6,850 & 12,116 & \checkmark& Oct, 2022\\
\bottomrule
\end{tabular}
}
\end{table}

\begin{figure}[tb]
 \centering
 \includegraphics[width=0.9\columnwidth,page=1]{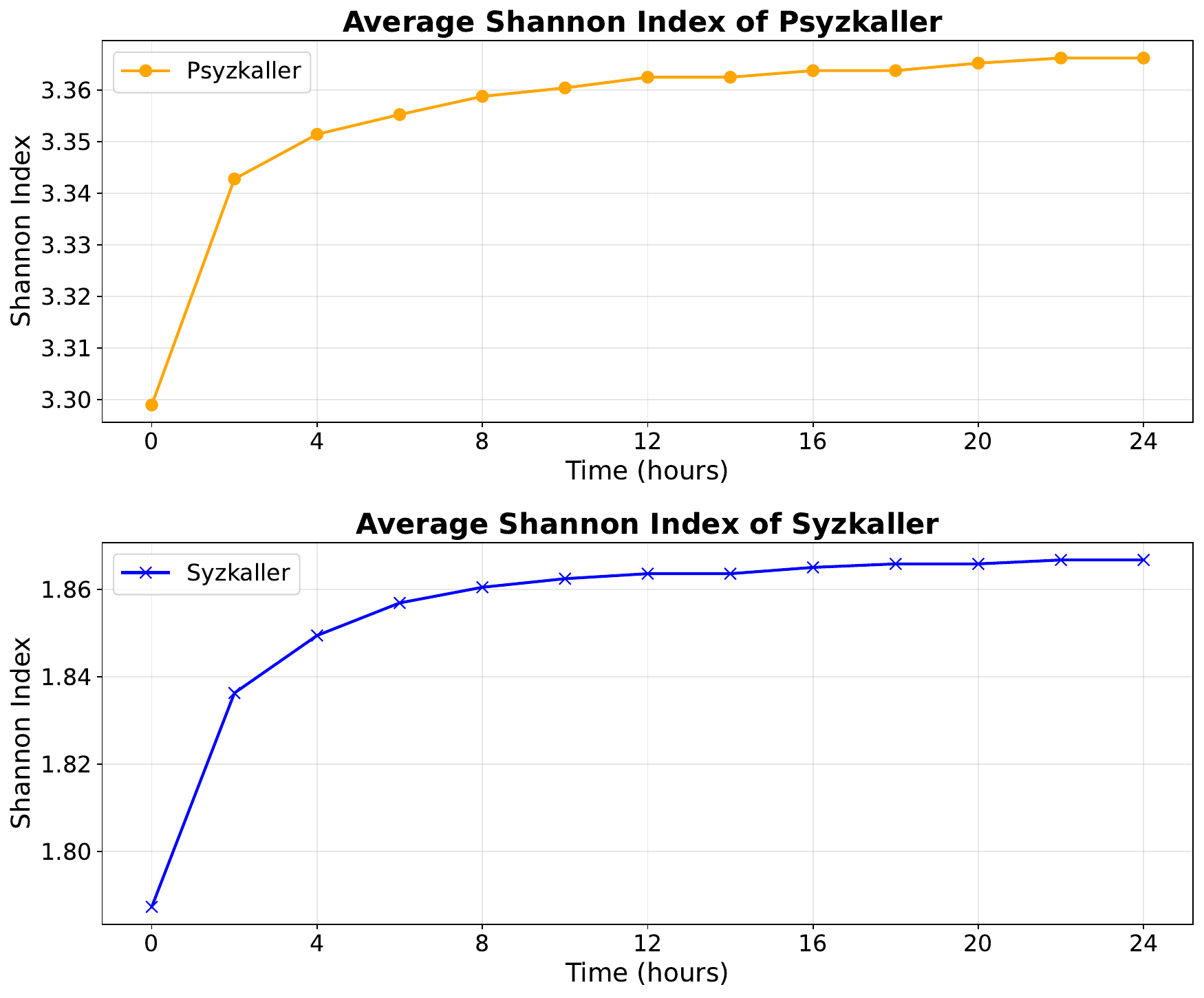}
 \caption{The diversity of $CT$ with and without extra SDR knowledge applied, by the metric of mean SI of all syscalls in $CT$. The original results (Syzkaller) are represented as the blue curve, and the results of fuzzing with enhanced SDR knowledge are illustrated as the yellow curve. }
 \label{fig:SI-expmt}
\end{figure}

Learning SDRs from historical and ongoing kernel execution data can have positive impact on enriching Syzkaller's Choice Table: it improves the diversity of choices for selecting the next syscall during seed generation. Figure~\ref{fig:SI-expmt} quantifies and visualizes this positive impact using the notion of Shannon Index (SI) from information theory~\cite{shannon1948}. 

Specifically, the Choice Table of Syzkaller, $CT$, is an $N \times N$ matrix, where $CT[i][j]$ specifies the probability of invoking syscall $s_j$ immediately after syscall $s_i$, and $N$ denotes the total number of syscalls supported by Syzkaller. 
The SI of $s_i$, denoted as $SI(s_i)$, is defined as $SI(s_i) = -\sum_{j=1}^{n} CT[i][j] \log_2 CT[i][j]$. A higher $SI(s_i)$ indicates greater diversity in the $i^{th}$ row of $CT$, implying that the probability mass is more evenly distributed and Syzkaller has a wider range of viable next syscalls to expand $s_i$.  

Figure~\ref{fig:SI-expmt} shows the evolution of the mean SI of $CT$ before (Syzkaller) and after integrating SDRs learned from DongTing and ongoing data (Psyzkaller)~\footnote{SDRs from normal and abnormal DongTing traces are equally weighted.} during 24-hour fuzzing on Linux-6.1. Psyzkaller increases the mean SI from approximately 1.83 to 3.30--3.50 bits. With low SI ($\sim$1.85 bits), Syzkaller exhibits a biased, low-uncertainty distribution over candidate syscalls when expanding seeds, restricting code-space exploration. In contrast, the higher SI ($>3.3$ bits) in Psyzkaller yields a flatter, higher-uncertainty distribution, where more candidates ($\sim$10 on average) have comparable selection probabilities. This enables broader exploration of execution paths and improves coverage.

\subsection{Integrating Trained Bigram with Syzkaller}
Each entry $CT[i][j]$ is the sum of two components: (1) a static value $CT_{st}[i][j]$, derived from syscall interfaces and their dependencies specified in syscall templates, and (2) a dynamic value $CT_{dyn}[i][j]$, which is periodically updated based on kernel feedback during fuzzing (e.g., newly discovered coverage or crashes attributable to the invocation of $s_i \rightarrow s_j$). Specifically,  Syzkaller updates $CT_{dyn}$ once every 33 new syscall sequences have been collected during fuzzing if the corpus of collected sequences is smaller than 100; or after every 333 new sequences being collected otherwise.

Accordingly, our approach employs two RPMs to learn SDRs from historical and on-the-fly data separately.  The first, denoted as $DTN$, is trained once using historical kernel execution data (i.e., the DongTing dataset) following  Algorithm~\ref{alg:learningSDR}. The second, denoted as $CorpusN$, is periodically updated during fuzzing using ongoing execution data.  Specifically, $CorpusN$ is updated in sync with Syzkaller updating $CT_{dyn}$ to avoid degrading fuzzing efficiency,  where all syscall sequences collected since the start of fuzzing are used to update $CorpusN$. 

To integrate these two RPMs into Syzkaller's workflow,  an Augmented Choice Table ($ACT$) is constructed to combine $CT$, $DTN$, and $CorpusN$. The construction and update procedure of  $ACT$ is as follows: at the start of fuzzing, each entry is initialized as $ACT[i][j] = CT_{st}[i][j]$.  Subsequently, whenever Syzkaller triggers an update of $CT_{dyn}$ upon enough new syscall sequences being collected, the corresponding $ACT$ entry is updated as $ACT[i][j]$ = $CT_{st}[i][j]$ + $CT_{dyn}[i][j]$ + $DTN[i][j]$ + $CorpusN[i][j]$.  Here, $DTN$ is deliberately excluded from the initialization step to avoid disrupting the complex startup procedures of Syzkaller. 

It should be noted that both $DTN$ and $CorpusN$ must undergo Syzkaller’s native transformation procedure—applying the square root to each entry and multiplying by two--—prior to their integration into $ACT$. This normalization prevents excessively large values in either matrix from dominating the seed generation process.

\subsection{The RandomWalk Stragtegy}
In Syzkaller, seeds (or SCSs) are generated uni-directionally: given a partially generated sequence $s_1 \cdot s_2 \ldots \cdot s_k$,  syscall $w_{k+1}$  is randomly selected from all syscalls $s_t$ such that $CT[k][t] > 0$ and appended to the end of the sequence.  We observe that this strictly forward expansion may limit the diversity of generated seeds and in turn the effectiveness of fuzzing.  To address this issue, we introduce a \emph{RandomWalk} strategy that enables bi-directional expansion of seeds: given $s_1 \cdot s_2 \ldots \cdot s_k$, expansion can proceed forward by appending a syscall $s_f$ to the end such that $ACT[s_k][s_f] > 0$, or backward by adding syscall  $s_b$  to the beginning of the sequence such that $ACT[s_b][s_1] > 0$.  The direction of expansion is chosen uniformly at random (i.e., each direction with probability 0.5).  

Algorithm~\ref{alg:genrandomwalk} implements the RandomWalk strategy. It begins by initializing a set of syscall pairs, $G$, with one syscall randomly selected from the set of all valid starting syscalls $S$. The algorithm then iteratively selects syscall  $(ap, \cdot)$ from $G$ to expand (Line~\ref{alg:rw:iter_start}). The expansion direction is determined in Lines~\ref{alg:rw:select_w}–\ref{alg:rw:select_dir}.  

For forward expansion, syscall $cf$ is first selected satisfying $DTN[ap][cf] > 0$ or $CorpusN[ap][cf] > 0$.  If no such syscall exists ($cf = -1$), $cf$ is chosen from candidates satisfying $ACT[ap][cf] > 0$.  The backward expansion is conducted analogously: syscall $cb$ is selected such that $DTN[cb][ap] > 0$ or $CorpusN[cb][ap] > 0$, with $ACT[cb][ap] > 0$ as a fallback. If this also fails, the algorithm switches to forward expansion.

Since Syzkaller guarantees that for every $ap \in CT$ there exists at least one $cf$ such that $CT[ap][cf] > 0$, Line~\ref{alg:rw:bw_ret_fw} always returns a valid candidate. Upon the success of forward expansion, the pair $(ap, cf)$ is added to $G$ as part of the generated sequence, and also stored in the $visited$ set to prevent redundant expansions. Backward expansions operate analogously, where $(cb, ap)$ is added to both $G$ and $visited$ upon success (Line~\ref{alg:rw:bw_succeed}).  

This iterative expansion continues until the constructed sequence reaches the length threshold in Syzkaller (typically 30). The final syscall sequence is then derived by performing a topological sort in $G$. Notably, the RandomWalk strategy in Algorithm~\ref{alg:genrandomwalk} performs not only on the starting and/or the end syscall of the target syscall sequences, but also any intermediate syscalls. This `generalization' may result in the pairs in $G$ ultimately forming a `syscall tree' rather than a one-dimensional syscall sequence.  However, the sorting algorithm ensures that the syscall sequences extracted from $G$ conform to the expanding order applied to the original sequence.

\begin{algorithm}
\caption{GenRandomWalk}
\label{alg:genrandomwalk}
\begin{algorithmic}[1]
\Require ACT: Augmented choice table; $L$: The length of syscall sequence to be generated; $S$: Set of starting syscalls.
\Ensure \textit{seq}, the generated syscall sequence.
\State $firstcall \gets$ {\sf RandomSelect}($S$)
\State $G \gets \{(-1, firstcall)\}$
\State $visited \gets \{(-1, firstcall)\}$

\While{{\sf Len}($G$) $<L$}\label{alg:rw:iter_start}
    \State $ap \gets$ {\sf RandomSelect}($G$)\label{alg:rw:select_w}
    \State $r \gets$ {\sf RandomSelect}($\{0,1\}$)\label{alg:rw:select_dir}
    \If{$r = 1$}
        \State $cb \gets$ {\sf ChooseBackwardRPM}($ap$)
        \If{$cb = -1$} \label{alg:rw:bw_fail_rpm}
            \State $cb \gets$ {\sf ChooseBackwardACT}($ap$)
        \EndIf
        \If{$cb = -1$}\label{alg:rw:bw_fail_act}
            \State $cf \gets$ {\sf ChooseFrontACT}($ap$)\label{alg:rw:bw_ret_fw}
            \If{$(ap,cf) \notin visited$}
                \State $visited \gets$ {\sf Append}(($ap,cf$), $visited$)
                \State $G \gets$ {\sf Append}(($ap,cf$), $G$)
            \EndIf
        \Else \label{alg:rw:bw_succeed}
            \If{$(cb, ap) \notin visited$}
                \State $visited \gets$ {\sf Append}($(cb, ap)$, $visited$)
                \State $G \gets$ {\sf Append}($(cb, ap)$, $G$)
            \EndIf
        \EndIf
    \Else \label{alg:rw:select_fw}
        \State $cf \gets$ {\sf ChooseFrontRPM}($ap$)
        \If{$cf = -1$}
            \State $cf \gets$ {\sf ChooseFrontACT}($ap$)
        \EndIf
        \If{$(ap, cf) \notin visited$}
            \State $visited \gets$ {\sf Append}($(ap, cf)$, $visited$)
            \State $G \gets$ {\sf Append}($(ap, cf)$, $G$)
        \EndIf
    \EndIf\label{alg:rw:end_fw}
\EndWhile
\State $seq \gets$ {\sf TopoSort}($G$)
\end{algorithmic}
\end{algorithm}

\section{Implementation}\label{sec:Implementation}

\begin{figure}[htbp]
	\centerline{\includegraphics[width=0.45\textwidth]{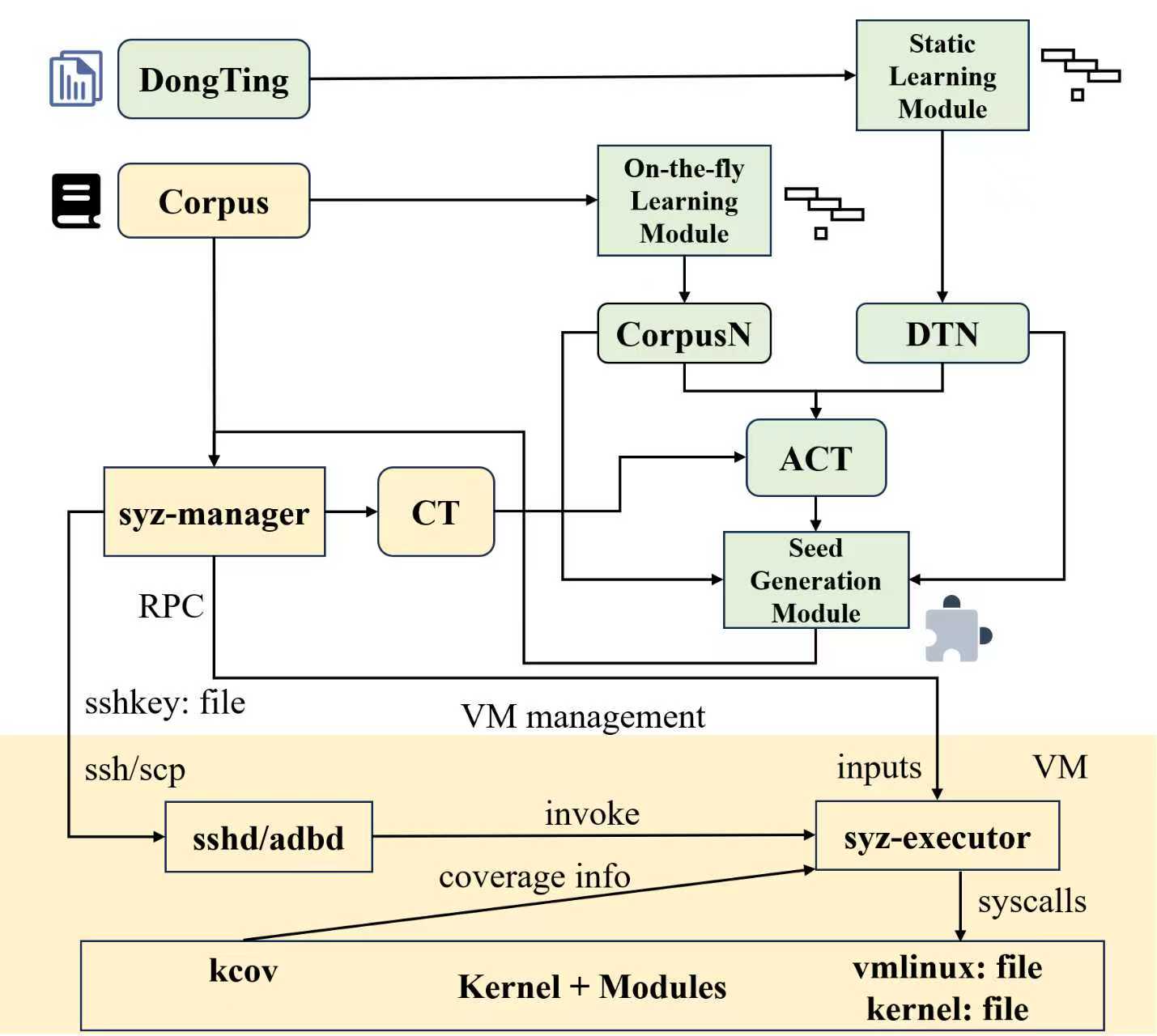}}
	\caption{System Architecture of Psyzkaller.}
	\label{fig:archi}
\end{figure}

We implement a prototype tool, called \textbf{Psyzkaller}, on top of Syzkaller. Figure~\ref{fig:archi} depicts the system architecture of Psyzkaller, where square components denote functional modules, rounded rectangles represent data structures, entities in yellow correspond to original Syzkaller components, and entities in green are introduced by Psyzkaller. 

Among these modules, the \texttt{Static Learning Module} is responsible for learning $DTN$, while the \texttt{On-the-fly Learning Module} incrementally updates $CorpusN$ using ongoing fuzzing results. The \texttt{Seed Generation Module} modifies Syzkaller’s seed construction procedures by incorporating both RPMs and the RandomWalk strategy (if enabled). Notably, all strategies—-learning from historical data, learning on-the-fly, and RandomWalk—-can be independently enabled, thereby facilitating the evaluation of their individual contributions to fuzzing performance. 

\subsection{Training with the DongTing Dataset}\label{ssec:preprocess}
Syzkaller distinguishes different syscall instances based on their parameter settings in the Choice Table. However, syscalls recorded in the DongTing dataset do not contain parameter information, as shown in the following excerpt: 
\begin{verbatim}
execve|brk|arch_prctl|access|openat|...
\end{verbatim}

Therefore, the DongTing traces must be pre-processed to restore syscall parameters before learning SDRs. To achieve this, we replicated the data-collection procedure described in~\cite{dongting} for all normal traces. Specifically, we executed the four Linux regression test suites on the same kernel version (Linux-5.15.0-153-generic) and collected syscall traces—including full parameter information—using \texttt{strace-4.18}.  The \texttt{trace2syz} tool~\cite{trace2syz} provided by MoonShine~\cite{pailoor_moonshine:_2018} was then applied to deserialize and convert the collected syscall sequences to the format acceptable by Syzkaller. This process produced a comparable number of normal traces (6,672) to DongTing’s original dataset (6,850), with all syscall parameters successfully restored.

Note that we excluded Linux kernel 5.15 from our experimental testbench to avoid interference and to better evaluate the usefulness of DongTing on other kernel versions, thereby demonstrating the generality of our method. 
%

%
For abnormal traces, the same pre-processing strategy cannot be applied, since these traces originated from PoC programs associated with reported crashes and vulnerabilities. Reproducing the PoCs to recover accurate syscall parameters is both complex and labor-intensive.  We therefore adopt an over-approximation strategy: for each syscall pair $s \rightarrow t$ appearing $n$ times in abnormal traces, we add $n$ to the counts of all instances $s'$, $t'$, and $s'\rightarrow t'$, where $s'$ and $t'$ are syscall instances in the Choice Table sharing the same syscall names as $s$ and $t$ but with different parameters. This approximation helps to ensure that potential SDRs relevant to vulnerabilities are not missed. 

Psyzkaller maintains separate RPMs trained on normal and abnormal traces, and combines them into a final RPM through weighted summation. Each RPM is scaled by a configurable weight factor before being aggregated.  For example, under a 1:2 weight ratio, the abnormal-trace RPM contributes twice as much as the influence of the normal-trace RPM. The combined matrix is then normalized so that each row sums to 1, ensuring a valid probability distribution. This mechanism enables us to control the relative influence of normal versus abnormal traces on the final Bigram model and, consequently on seed generation.

\section{Evaluation}\label{sec:Evaluation}
We conducted a series of experiments to evaluate the effectiveness and efficiency of \texttt{Psyzkaller}. These experiments aim to answer the following research questions.
\begin{itemize}
  \item \textbf{RQ1}: How do different strategies for learning from DongTing influence the performance of Psyzkaller?  
  \item \textbf{RQ2}: How `good’ are the SDRs learned from historical and ongoing fuzzing data?
  \item \textbf{RQ3}: How do individual strategies in Psyzkaller affect the fuzzing performance, and which combination yields the best overall results?  
  \item \textbf{RQ4}: How does the fuzzing performance of Psyzkaller compared to that of SOTA fuzzers?  
\end{itemize}

\textbf{Testbench:}  four Linux kernel versions were selected as fuzzing objects, namely 5.19 (widely deployed stable release), 6.1 (long-term support release), 6.2 (widely studied by other fuzzers), and 6.12 (the latest at the time of experimentation), where version 6.2 was only considered when comparing Psyzkaller with other fuzzers.  Each fuzzing campaign was repeated three times per kernel version, with each run lasting 48 hours.  Reported results are averaged across repetitions to reduce the impact of randomness of fuzzing outcomes.
%


{\bf Comparison objects:} we compare the performance of Psyzkaller against Syzkaller,  ACTOR~\cite{actor}, and SyzDescribe~\cite{syzdescribe}, which are representatives of leveraging dynamic or static SDR learning to improve fuzzing effectiveness.  

\subsection{Evaluating Training Strategies on the DongTing Dataset}\label{ssec:eval_dongting}

We evaluate the impact of different learning strategies on the DongTing dataset by applying Psyzkaller to selected Linux kernel versions under varying learning weight ratios. To isolate the effect of the training strategy, we disable RandomWalk and on-the-fly learning during fuzzing.

\textbf{Using different weight ratios to learn from normal and abnormal traces in DongTing}.  Although infinitely many weight ratios are possible, our experiments evaluated five representative configurations: 1:1 (denoted as MIX11), 1:2 (MIX12), 2:1 (MIX21),  abnormal traces only (ABN), and normal traces only (NOR). These settings correspond to assigning equal influence to SDRs learned from normal and abnormal traces, or emphasizing one category over the other.

\begin{figure}[htbp]
    \centerline{\includegraphics[width=0.9\columnwidth,page=1]{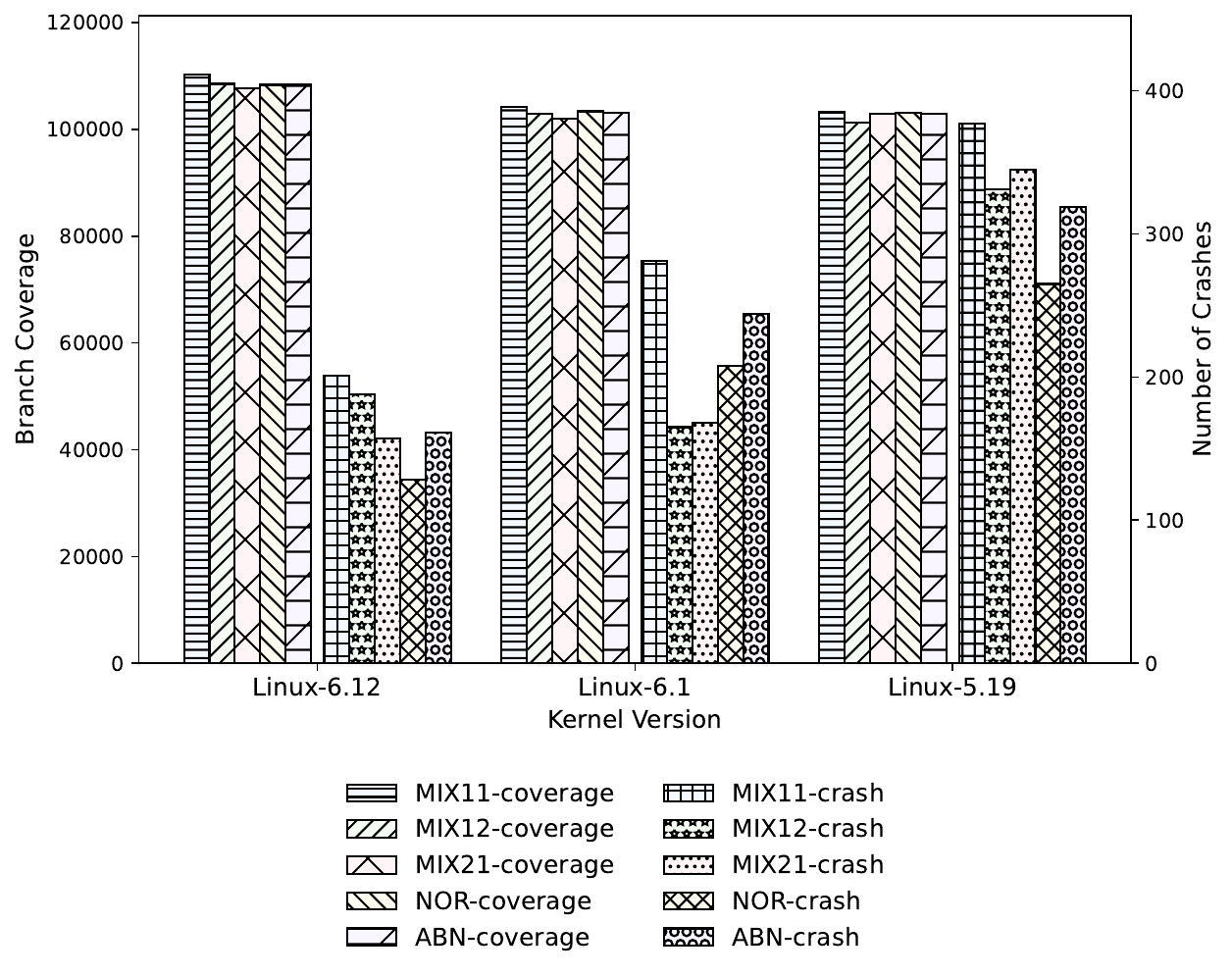}}
    \caption{Branch Coverage and Number of Crashes of Psyzkaller under Different Learning Weight Ratios}
    \label{fig:ratio}
\end{figure}

Figure~\ref{fig:ratio} shows the branch coverage and number of crashes achieved by Psyzkaller under each learning ratio.  Due to space constraints, both metrics are illustrated in a single figure:  the left y-axis represents branch coverage, while the right y-axis represents crash counts. Solid bars indicate branch coverage results, while hatched bars denote crash counts. This convention is used throughout this section unless otherwise noted.  

As illustrated, MIX11—weighting normal and abnormal traces equally—achieves the best overall performance across both branch coverage and crash detection. This indicates that a balanced combination of learning SDRs from normal traces and the vulnerability-relevant SDRs from abnormal traces yields the most effective historical-model training. Accordingly, we adopt the 1:1 ratio in all subsequent experiments when training Psyzkaller on the DongTing dataset.

\textbf{Learning from Part of DongTing}. It is important to understand how the effectiveness of our approach degrades when the historical dataset used to train the Bigram model becomes smaller or less comprehensive. To this end, we conduct experiments using randomly sampled subsets of the DongTing traces at 100\%, 80\%, 50\%, and 20\% of the full dataset. We focus on DongTing due to the limitations of alternative datasets discussed earlier.

\begin{figure}[htbp]
    \centerline{\includegraphics[width=0.9\columnwidth,page=1]{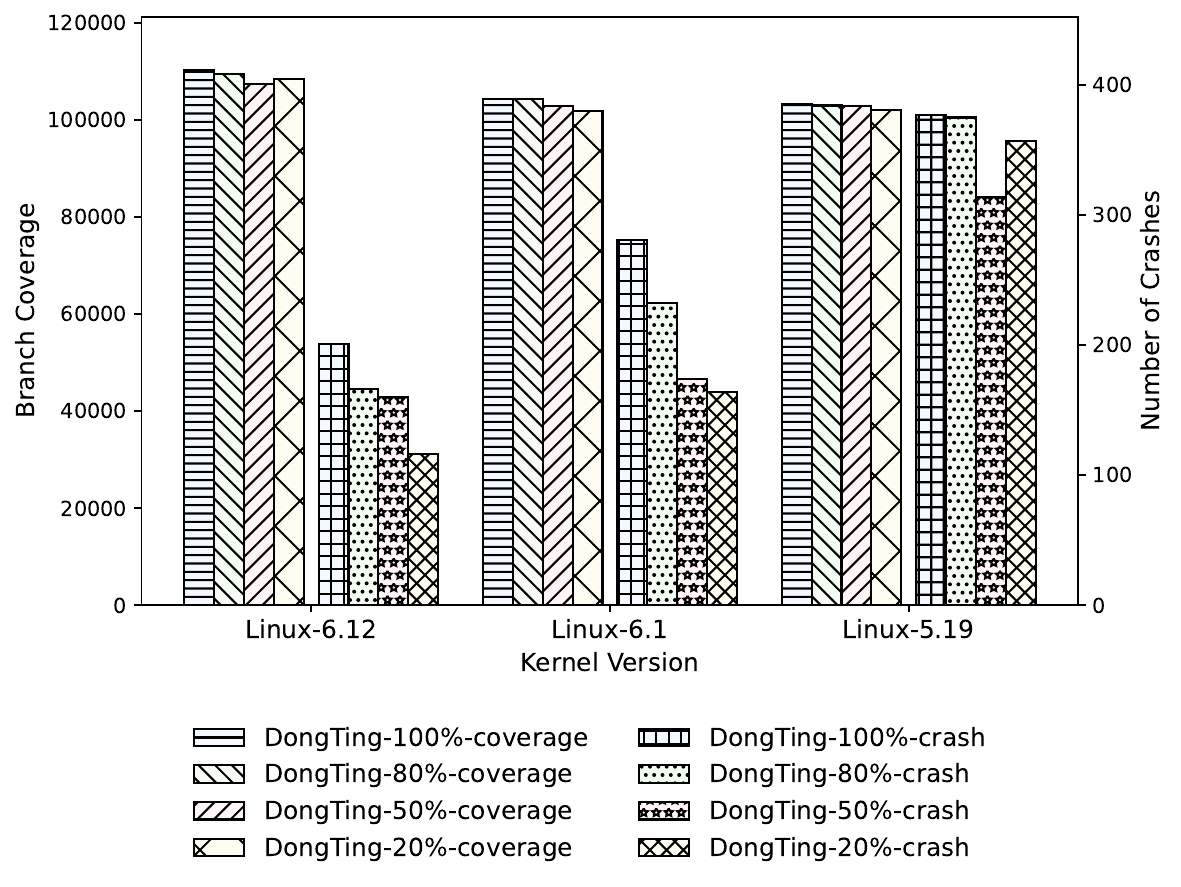}}
    \caption{Branch Coverage and Number of Crashes of Psyzkaller when Trained on Different DongTing Subsets}
    \label{fig:size}
\end{figure}

Figure~\ref{fig:size} presents the fuzzing performance of Psyzkaller when trained on these different subsets of DongTing.  As expected, the best performance is achieved when the full dataset is used.  Overall, fuzzing performance (both branch coverage and crash count) declines as the size of the training data decreases. Two exceptions were observed: DongTing-20\% ranked the second lowest in performance twice across all experiments. We suspect this is due to variations in the quality of the randomly sampled traces, particularly regarding their ability to expose meaningful SDRs.

These results indicate that larger and more comprehensive datasets generally enable Psyzkaller to learn more accurate and complete SDRs, which in turn leads to improved fuzzing performance. At the same time, small historical datasets can still contribute to improving fuzzing performance, as described in subsequent experiments. 

\textbf{Answer to RQ1:} Giving equal weights to SDRs learned from normal and abnormal traces out of the full Dongting dataset helps Psyzkaller to achieve the best fuzzing performance. 

\subsection{Genuineness of Learned SDRs}\label{ssec:eval_sdr}
Bigram-based learning captures consecutive syscall co-occurrences as a proxy for SDRs, but inevitably introduces spurious relations. For instance, the relation $getsockopt \rightarrow{ } openat$ is identified during our experiments but is spurious, as the two syscalls operate on unrelated resources (socket options vs.\ files) in different kernel subsystems. This raises a key question: \texttt{how genuine are the SDRs learned from historical and ongoing fuzzing data?}

To answer this, we quantitatively evaluate SDR genuineness using models trained on the DongTing dataset (best 1:1 weighting) and ongoing fuzzing results from Linux kernel v6.12. For each trained Bigram model (DTN or CorpusN), we randomly sample 20 rows and select the top-20 SDRs per row with the largest weight gains over Syzkaller, yielding 400 SDRs per setting. Genuineness is assessed via LLM-assisted analysis followed by manual validation: each SDR is evaluated by ChatGPT~5 and Claude~4.6 using syscall documentation and few-shot examples of genuine vs.\ spurious relations, and disagreements are resolved by the authors.

Results show 317/400 (79.25\%) SDRs learned from ongoing fuzzing and 338/400 (84.50\%) from DongTing are genuine. Across the 655 validated SDRs, weights increase over Syzkaller’s Choice Table by 41–11,258 (avg.\ 1,373.29), with 393 previously assigned zero weight. These findings indicate our approach not only discovers novel genuine SDRs but also substantially improves their prioritization for seed generation.

\textbf{Answer to RQ2:} SDRs learned by Psyzkaller from both historical and ongoing fuzzing data exhibit high validity.

\subsection{Impacts of Individual Strategies}\label{ssec:eval_str}
Psyzkaller is designed so that individual strategies can be enabled or disabled, allowing us to isolate and evaluate their individual and collective impact on fuzzing performance. These strategies include: learning SDRs from the DongTing dataset (denoted as D), learning SDRs on-the-fly from the corpus of minimized syscall sequences collected during fuzzing (denoted as C), and the RandomWalk strategy (denoted as R).  When all strategies are disabled, Psyzkaller effectively reduces to vanilla Syzkaller. 

\begin{figure}[htbp]
   \centerline{\includegraphics[width=0.9\columnwidth]{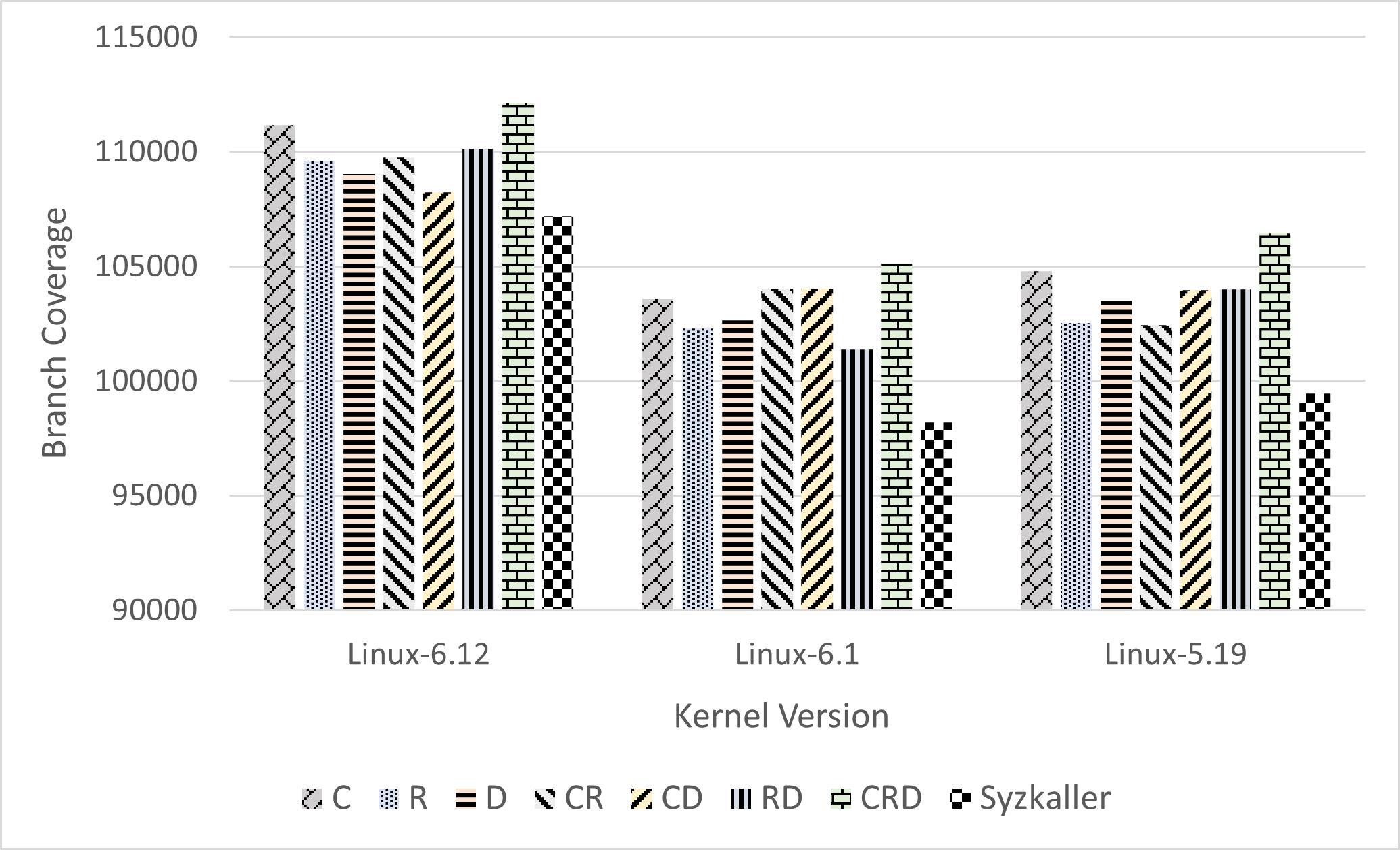}}
    \caption{Branch Coverage Achieved by Psyzkaller under Different Strategy Settings}
    \label{fig:cov-str}
\end{figure}

\begin{figure}[htbp]
    \centerline{\includegraphics[width=0.9\columnwidth]{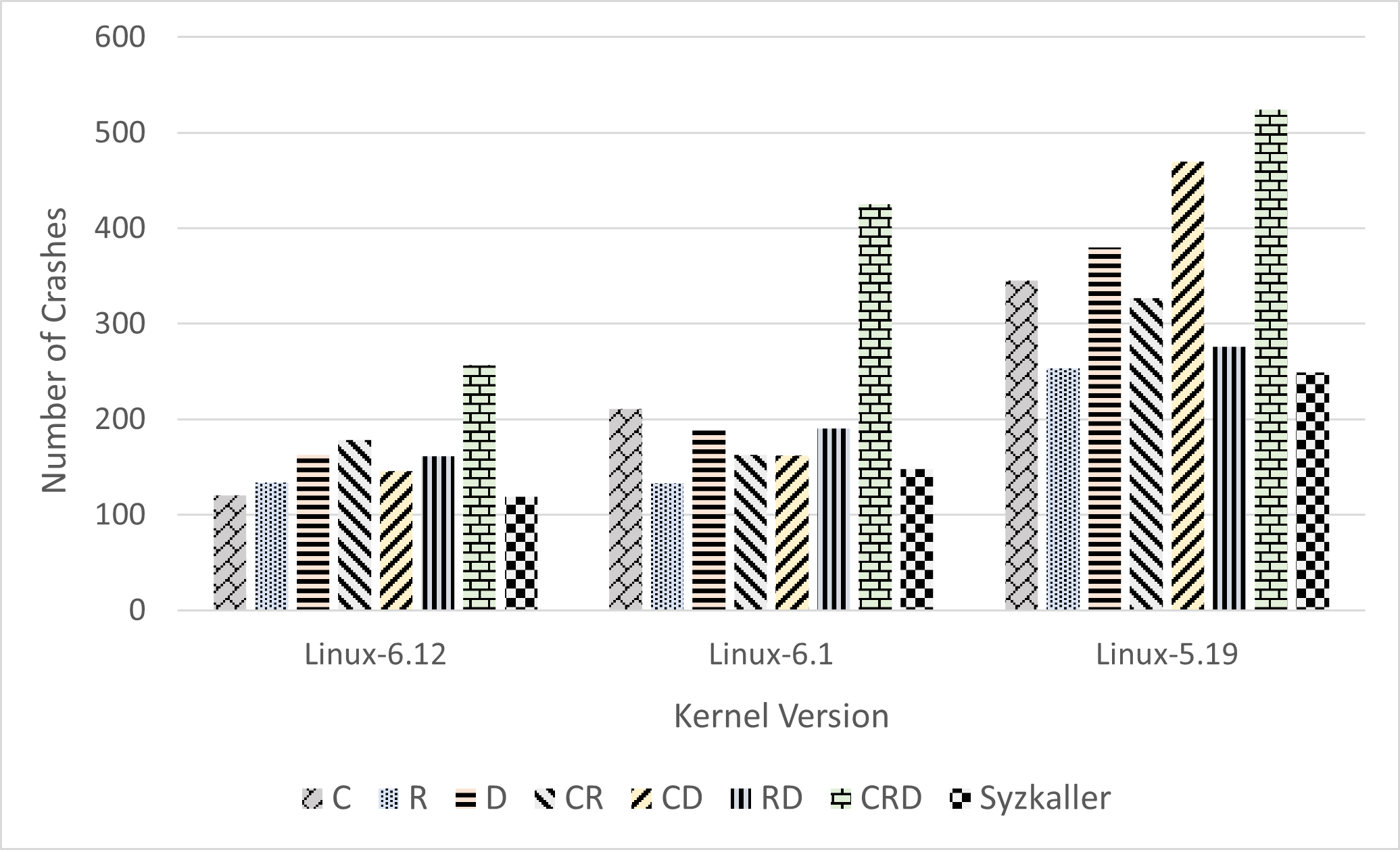}}
  \caption{Number of Crashes Detected by Psyzkaller under Different Strategy Settings}
 \label{fig:crash-str}
\end{figure}

We evaluated Psyzkaller on the testbench under different combinations of these strategies, while vanilla Syzkaller was also applied as a baseline. Figure~\ref{fig:cov-str} shows the branch coverage achieved under each strategy configuration, while Figure~\ref{fig:crash-str} presents the corresponding number of detected crashes.

Several key observations can be drawn from Figures~\ref{fig:cov-str} and \ref{fig:crash-str}:

\begin{itemize}
\item Psyzkaller outperformed Syzkaller across all kernel versions whenever at least one strategy was enabled. The only exception occurred in kernel version 6.1, where the RandomWalk strategy alone resulted in fewer crashes than Syzkaller. This is likely because RandomWalk introduces additional randomness, which may divert fuzzing from critical code paths to vulnerabilities while still enhancing exploration of previously unseen code regions.

\item The best overall fuzzing performance—both in branch coverage and number of crashes—was consistently achieved when all three strategies were enabled (RCD), regardless of kernel version.

\item Training the Bigram model with on-the-fly fuzzing results (C) consistently yielded higher branch coverage than training with the DongTing dataset (D). Conversely, training with DongTing led to more crashes being detected in two of the three kernel versions. This suggests that on-the-fly traces better reflect the execution space of the current kernel, while the DongTing dataset, despite being collected from older kernels, still provides valuable abnormal traces for vulnerability discovery.  

\item Combining two strategies does not always outperform using each strategy independently, and the effects vary by kernel version. For example, combining C and D (CD) produced lower branch coverage than C alone in two kernel versions. This phenomenon is particularly pronounced for RandomWalk. Augmenting learning from DongTing with it (RD) increases branch coverage compared to D alone in two evaluated kernel versions, but reduces crash counts in two versions. Meanwhile, combining RandomWalk with on-the-fly learning (RC) improves both branch coverage and crash counts over C in only one kernel version.
\end{itemize}

\textbf{Interaction between RandomWalk and Other Strategies.} To confirm this `conflict', we evaluate the impact of RandomWalk on seed quality on Linux kernel v6.12 using average seed length and effective test ratio (i.e., tests passing kernel validity checks) as metrics. As shown in Table~\ref{tab:randomwalk_seed_quality}, combining RandomWalk with a single learning strategy (RC or RD) degrades seed quality compared to C or D alone, yielding shorter seeds and lower effective test ratios. In contrast, enabling both strategies (RCD) significantly improves quality, increasing the effective test ratio from 86.38\% to 90.52\%. These results show that RandomWalk may conflict with individual learning strategies, but this effect is reversed when C and D are jointly applied.

\begin{table}[ht]
\centering
\caption{Impact of RandomWalk on Seed Quality}
\begin{tabular}{ccc}
\hline
Optimizations & Avg. Seed Len. & Eff. Test Ratio (\%) \\
\hline
C vs. RC  & 30.00 vs. 28.43 & 83.02 vs. 80.39 \\
\hline
D vs. RD & 30.00 vs. 29.25  & 84.89 vs. 85.41 \\
\hline
CD vs. RCD & 30.00 vs. 29.23 & 86.38 vs. 90.52 \\
\hline
\end{tabular}
\label{tab:randomwalk_seed_quality}
\end{table}

To further investigate this, we vary the consultation priority in RandomWalk and evaluate its impact on fuzzing performance. We consider several strategies: always consulting the Choice Table (CT-only), prioritizing Bigram models trained on historical data (DTN-first), ongoing fuzzing data (CorpusN-first), or both (Combined-first), as well as randomly selecting among these models (Random). If the prioritized source fails, the strategy falls back to the Choice Table. We also evaluate variants where the fallback is another Bigram model (DTN-first+C and CorpusN-first+D) with the Choice Table. 

Table~\ref{tab:randomwalk_pri} shows consultation priority significantly affects performance. CT-only yields the worst results, underscoring the importance of SDRs. Randomized consultation performs the second worst, due to inconsistent guidance. Prioritizing a single Bigram model improves performance but remains suboptimal, indicating partial SDR knowledge may still conflict with RandomWalk. Adding a second learning source in the fallback (i.e., DTN-first+C or CorpusN-first+D) consistently improves results, suggesting richer SDR knowledge mitigates such conflicts. Prioritizing the combined model (Combined-first) achieves the best coverage and crash detection.

Overall, these results show that the effectiveness of RandomWalk depends on the sufficiency of SDR knowledge: more comprehensive guidance leads to better integration. Notably, with both learning strategies enabled, RandomWalk consistently yields positive gains regardless of consultation priority.

\begin{table}[ht] 
\centering \caption{Effect of Different Consult Priorities in RandomWalk on Learning Stratgies}
\begin{tabular}{lll} \hline Priorties & Coverage & Crash Num. \\ \hline 
CT-only & $107,664_{(-2,584)}$ & $117_{(-128)}$ \\ 
CorpusN-first & $108,009_{(-2,239)}$ & $138_{(-107)}$\\ 
CorpusN-first+D & $108,982_{(-1,266)}$ & $165_{(-80)}$ \\ 
DTN-first & $108,770_{(-1,478)}$ & $182_{(-63)}$ \\ 
DTN-first+C & $109,546_{(-702)}$ & $200_{(-45)}$ \\ 
Random & $107,790_{(-2,458)}$ & $134_{(-111)}$ \\ 
Combined-first (=RCD) & $110,248_{(0)}$ & $245_{(0)}$ \\ 
\hline \end{tabular} \label{tab:randomwalk_pri} \end{table}

\textbf{Answer to RQ3:} Each strategy in Psyzkaller improves fuzzing performance on the tested kernels compared to Syzkaller, and enabling all three strategies achieves the best overall results.

\subsection{Comparison with SOTA Fuzzers}\label{ssec:eval_fuzzer}
When comparing against SOTA fuzzers, we evaluate Psyzkaller against Syzkaller (the baseline) across three kernel versions in our testbed (results shown in Figures~\ref{fig:cov-str} and~\ref{fig:crash-str}). We further compare Psyzkaller, with RCD enabled, against Syzkaller, ACTOR, and SyzDescribe on Linux kernel version 6.2. We include ACTOR and SyzDescribe because they are the only approaches that, like Psyzkaller, improve fuzzing by inferring SDRs and that we were able to successfully rebuild. More broadly, although many kernel fuzzing techniques have been open-sourced, most cannot be reliably reproduced across arbitrary kernel versions due to complex software dependencies and incomplete implementation details.

\textbf{Comparison with Syzkaller.} Table~\ref{tab:pvs} summarizes the branch coverage and number of crashes achieved by Syzkaller and Psyzkaller across the three kernel versions. It is evident that Psyzkaller consistently outperformed Syzkaller. Specifically, Psyzkaller improved branch coverage by 4.6\%–7.0\% depending on the kernel version. In terms of crash detection, Psyzkaller identified 110.4\%–187.2\% more crashes than Syzkaller, detecting a total of 1,206 crashes compared to 516 by Syzkaller.

\begin{table}
\centering
\caption{Performance of Psyzkaller vs. Syzkaller}
\label{tab:pvs}
\begin{tabular}{lllll}
  \toprule
  Kernel & Metric & Syzkaller & Psyzkaller \\
  \midrule
  \multirow{2}{*}{5.19} & \# of Crashes & 249    & $524_{(+110.4\%)}$ \\
                        & \# of Cov.    & 99,469  & $106,435_{(+7.0\%)}$ \\
  \midrule
  \multirow{2}{*}{6.1}  & \# of Crashes & 148    & $425_{(+187.2\%)}$ \\
                        & \# of Cov.    & 98,191  & $105,113_{(+7.0\%)}$ \\
  \midrule
  \multirow{2}{*}{6.12} & \# of Crashes & 119    & $257_{(+116.0\%)}$ \\
                        & \# of Cov.    & 107,173 & $112,125_{(+4.6\%)}$ \\
  \bottomrule
\end{tabular}
\end{table}

\textbf{Comparison with SOTA fuzzers.} ACTOR is a Syzkaller-based fuzzer that learns syscall dependencies associated with certain types of vulnerabilities. SyzDescribe is a syscall template generator for Syzkaller, specifically targeting device driver-related syscalls. Both tools support only a limited range of kernel versions, with version 6.2 being the only version supported by both. Therefore, we use kernel version 6.2 as the target for this comparison.

Psyzkaller and ACTOR both leverage discovered SDRs to guide seed generation, making them difficult to combine. In contrast, SyzDescribe can be integrated with Psyzkaller, since our extensions do not modify Syzkaller’s syscall templates. To evaluate the combined effect of multiple optimization strategies, we created a hybrid tool by integrating SyzDescribe’s templates with Psyzkaller, denoted as PsyzD.  Accordingly, we compare Psyzkaller against four fuzzers: ACTOR, SyzDescribe, Syzkaller, and PsyzD.

Table~\ref{tab:pvsota} summarizes the comparison results. Columns 2 and 4 report the absolute branch coverage and crash counts achieved by each fuzzer, respectively, while Columns 3 and 5 present the corresponding differences relative to Psyzkaller (negative values indicate performance degradation).

Psyzkaller achieves the highest branch coverage, exceeding ACTOR by 4.1\% (4,176 / 101,970), SyzDescribe by 6.6\% (6,609 / 99,537), and Syzkaller by 5.4\% (5,410 / 100,736).  PsyzD achieves the second-highest coverage, surpassing SyzDescribe.  A similar pattern is observed for crash detection: Psyzkaller and PsyzD discover the largest and second-largest numbers of kernel crashes, respectively. Specifically, Psyzkaller detects 151.9\% more crashes than ACTOR (240 vs. 158), 57.9\% more than SyzDescribe (146 vs. 252), and 75.3\% more than Syzkaller (171 vs. 227).

Another observation is that ACTOR detected fewer crashes and SyzDescribe achieved lower branch coverage compared to Syzkaller. This is likely because both tools were released two years ago, while Syzkaller has continuously integrated optimizations developed over the past two years, such as KernelGPT~\cite{yang_kernelgpt_2025}.



\begin{table}[hb]
\centering
\caption{Fuzzing Performance on Kernel v6.2}
\label{tab:pvsota}
\begin{tabular}{lllll}
  \toprule
   Fuzzers & Branch Coverage & Number of Crashes \\
  \midrule
  Psyzkaller & $106,146_{(0)}$ & $398_{(0)}$\\
  ACTOR & $101,970_{(-4,176)}$ & $158_{(-240)}$ \\
  SyzDescribe & $99,537_{(-6,609)}$ & $252_{(-146)}$ \\
  PsyzD & $102,508_{(-3,638)}$ & $286_{(-112)}$ \\
  Syzkaller & $100,736_{(-5,410)}$ & $227_{(-171)}$ \\
  \bottomrule
\end{tabular}
\end{table}



It is noteworthy that PsyzD achieves better performance than SyzDescribe alone, but still underperforms compared to Psyzkaller. We argue that, while the syscall templates generated by SyzDescribe enhance Psyzkaller’s ability to explore device driver code, they may also divert attention from other kernel modules, resulting in a trade-off in overall performance. Specifically, in the DongTing dataset, SDRs related to device drivers constitute only a small portion of the full $DTN$—roughly less than 10\%.  By increasing the likelihood of generating device driver-related syscall sequences, SyzDescribe’s templates may inadvertently reduce exploration of other kernel components that are easier to expand coverage and trigger crashes.

To investigate this hypothesis, we conducted additional experiments focusing exclusively on fuzzing device drivers with Psyzkaller, PsyzD, and SyzDescribe.  We modified Syzkaller’s configuration to exclude syscalls unrelated to device drivers. The resulting branch coverage and crash counts are summarized in Table~\ref{tab:pvsota_dr}. From the table, two observations emerge when targeting device drivers: (1) PsyzD surpasses Psyzkaller in both branch coverage and crash detection; (2) SyzDescribe achieves the highest branch coverage but detects the fewest crashes.

The first observation supports our hypothesis.  Regarding the second observation, two factors might contribute: (i) SyzDescribe’s exploration of device driver code is partially undermined by Psyzkaller’s RandomWalk strategy, which explores kernel space more broadly but without focus;  (ii) SyzDescribe lacks information from previous crashes.  Consequently, with the inclusion of abnormal sequences from the DongTing dataset, both Psyzkaller and PsyzD gain greater capability in triggering crashes.



\begin{table}[htbp]
\centering
\caption{Fuzzing Performance on Device Drivers (v6.2) }
\label{tab:pvsota_dr}
\begin{tabular}{lllll}
  \toprule
  Fuzzers & Branch coverage & Number of Crashes \\
  \midrule
  Psyzkaller & $72,288_{(0)}$ & $243_{(0)}$\\
  SyzDescribe & $72,855_{(+567)}$ & $227_{(-16)}$ \\
  PsyzD & $72,373_{(+85)}$  & $265_{(+22)}$ \\
  \bottomrule
\end{tabular}
\end{table}

\textbf{Comparison on Throughput.} We further compare the throughput of Syzkaller and Psyzkaller under a fixed 48-hour budget, measured by the number of executions. As shown in Table~\ref{tab:throughput}, Psyzkaller achieves throughput comparable to Syzkaller across all four kernel versions, with differences ranging from -1.51\% to 1.25\%, and slightly higher throughput in three cases.

This parity stems from two factors. First, Psyzkaller generates seeds of similar length to Syzkaller (29.30 vs.\ 30 on average). Since Syzkaller caps seed length at 30 and most seeds reach this limit, Psyzkaller adopts the same constraint (to ensure fair comparison), resulting in comparable execution time per test case. Second, although Psyzkaller incurs higher seed generation overhead (194.1\%--643.0\%), it has negligible impact on throughput because seed generation and execution are pipelined in Syzkaller’s architecture, effectively hiding the additional cost.

\begin{table}
\centering
\caption{Throughput of Psyzkaller vs. Syzkaller}
\label{tab:throughput}
\begin{tabular}{lllll}
  \toprule
  Kernel & Metric & Syzkaller & Psyzkaller\\
  \midrule
  \multirow{3}{*}{5.19} 
    & Exec Total          & 4,451,942 & $4,481,946_{(+0.67\%)}$ \\
    & Seed Len    & 30.00   & $29.30_{(-2.33\%)}$ \\
    & Gen Time (ms) & 3.281   & $21.952_{(+569.1\%)}$ \\
  \midrule
  \multirow{3}{*}{6.1}  
    & Exec Total          & 5,659,660 & $5,720,021_{(+1.07\%)}$ \\
    & Seed Len    & 30.00   & $29.32_{(-2.27\%)}$ \\
    & Gen Time (ms) & 5.870   & $17.263_{(+194.1\%)}$\\
  \midrule
  \multirow{3}{*}{6.2}  
    & Exec Total          & 4,054,188 & $3,992,911_{(-1.51\%)}$ \\
    & Seed Len    & 30.00   & $29.33_{(-2.23\%)}$ \\
    & Gen Time (ms) & 2.657   & $18.718_{(+604.5\%)}$ \\
  \midrule
  \multirow{3}{*}{6.12} 
    & Exec Total          & 4,481,169 & $4,537,233_{(+1.25\%)}$ \\
    & Seed Len    & 30.00   & $29.32_{(-2.27\%)}$ \\
    & Gen Time (ms) & 3.476   & $21.086_{(+506.6\%)}$ \\
  \bottomrule
\end{tabular}
\end{table}

\subsubsection{Crash Triage}\label{sssec:vul}
We collected all of the 1,206 crashes detected in the experiments to analyze their root causes and identify previously unknown vulnerabilities, with a goal of comparing the vulnerability detection capabilities of Psyzkaller and SOTA tools.
%
%
The analysis began with using Syzkaller’s built-in \texttt{syz-repro} tool~\cite{syz-repro} to generate PoCs for the detected crashes. This step yielded a total of 426 PoCs, with 355 generated from the 1,206 crashes detected by Psyzkaller and 70 from the 516 crashes detected by Syzkaller.  

We then conducted a thorough investigation to determine whether these PoCs had been previously reported, consulting Syzbot (Syzkaller’s official crash reporting service)~\cite{syzbot}, the Syzkaller community mailing lists~\cite{syzkallercommunity}, and the CVE database~\cite{cvedetails}. By analyzing both the generated PoCs and their associated crash logs, we identified 22 previously unreported PoCs detected by Psyzkaller and 1 from Syzkaller.

To further validate these potentially new vulnerabilities, we executed the unreleased 23 PoCs in a computer running Ubuntu 25.04 (Linux-6.14). This step verified these PoCs' validity in the most recent operating systems.
It confirmed that all the 23 PoCs successfully triggered the associated crashes. 


Lastly, we investigated the source codes of the 23 PoCs for root cause analysis. This analysis identified eight (8) previously unknown vulnerabilities discovered by Psyzkaller~\footnote{A single vulnerability may manifest as multiple crashes, resulting in multiple reproducible PoCs}, and one (1) found by Syzkaller. Table~\ref{tab:newvuln} summarizes these identified vulnerabilities. Note that all identified vulnerabilities have been reported to the China National Vulnerability Database (CNVD), and their public disclosure has been deferred pending confirmation or rejection by CNVD, in accordance with responsible disclosure practices.

\begin{table*}[htbp]
\centering
\caption{Previously Unknown Vulnerabilities Detected by Psyzkaller and Syzkaller}
\label{tab:newvuln}
\begin{tabular}{ccccc}
\toprule
\textbf{Kernel Version} & \textbf{Type} & \textbf{Function Affected} & \textbf{Description} & \textbf{Detected by} \\
\midrule
Linux-6.12 & DEADLOCK & loop\_reconfigure\_limits & Triple Deadlock & Psyzkaller \\
Linux-6.12 & DEADLOCK & loop\_reconfigure\_limits & Triple Deadlock & Psyzkaller \\
Linux-6.12 & DEADLOCK & \_\_submit\_bio & Triple Deadlock & Psyzkaller \\
Linux-6.12 & DEADLOCK & \_\_submit\_bio & Triple Deadlock & Psyzkaller \\
\midrule
Linux-6.1 & DEADLOCK & ext4\_multi\_mount\_protect & Deadlock by Locking & Psyzkaller \\
Linux-6.1 & DEADLOCK & ext4\_map\_blocks & Deadlock by Locking & Psyzkaller \\
Linux-6.1 & WARNING & cipso\_v4\_sock\_setattr & Violating RCU Specification & Syzkaller \\
\midrule
Linux-5.19 & WARNING & sel\_write\_load & May Zero-Size Vmalloc & Psyzkaller \\
Linux-5.19 & WARNING & reserve\_ds\_buffers & Memory Allocation May Fail & Psyzkaller \\
\bottomrule
\end{tabular}
\end{table*}

Notably, six vulnerabilities found by Psyzkaller are deadlock-related, highlighting its ability to leverage learned SDRs to generate complex syscall sequences that expose unsafe cross-subsystem interactions. For example, a triple-deadlock in \texttt{loop\_reconfigure\_limits} is triggered by three threads (A, B, C), each holding a lock (\texttt{q\_usage\_counter}, \texttt{q->limits\_lock}, and \texttt{xattr\_sem}) while requesting another in a conflicting order: A holds \texttt{q\_usage\_counter} and requests \texttt{q->limits\_lock}; B holds \texttt{q->limits\_lock} and requests \texttt{xattr\_sem}; C holds \texttt{xattr\_sem} and requests \texttt{q\_usage\_counter}, forming a cycle.

Constructing such a PoC requires capturing intricate cross-module syscall dependencies, which is highly challenging. We attribute Psyzkaller’s effectiveness to SDRs learned from large-scale kernel execution data: normal traces (DongTing) capture dependencies for deep exploration, while abnormal traces bias toward vulnerability-triggering patterns. Combined with on-the-fly learning, Psyzkaller continuously refines SDRs to explore unseen paths. Notably, DongTing’s abnormal traces originate from \texttt{syzbot} and similar sources, yet none covers the eight new vulnerabilities discovered by Psyzkaller.


\textbf{Answer to RQ4:} In its optimal strategy configuration, Psyzkaller outperforms Syzkaller, ACTOR, and SyzDescribe in both branch coverage and vulnerability detection. During the experiments, Psyzkaller demonstrates comparable througput as Syzkaller.

\section{Discussion}\label{sec:Discussion}

{\bf SOTA Fuzzers for Comparison}. We have devoted extensive efforts to including SOTA fuzzers in our evaluation; however, practical challenges precluded their inclusion. Specifically, several tools lack adequate documentation on their operational requirements. For instance, SyzVages~\cite{syzvegas-git} provided no explicit specifications of environment dependencies in its documentation, with certain required libraries no longer available. Mock~\cite{mock-git} offered a brief usage overview that lacks the detail of critical environment dependencies, preventing it from being compiled successfully due to unresolvable library version conflicts. Similar library conflicts occurred to SyzGen++~\cite{syzgen++-git}.  Other tools removed several critical components of their optimization techniques for copyright or unclear reasons, such as Healer~\cite{sun_healer_2021}, whose publicly-available version exhibited a performance that is markedly below the results reported in its research paper.  Consequently, these tools could not be compiled or executed for comparative evaluation.

{\bf Crash filtering}.  Among the crashes reported by Syzkaller, a subset are explicitly classified by the tool as non-kernel-related.  Specifically, these crashes either stem from tool-specific anomalies (e.g., those labeled ``SYZFAIL") or have undetermined root causes (e.g., the ``no output from test machine" error). Since such non-kernel-related issues introduce noise into statistical and root cause analyses of kernel crashes, they were filtered out based on Syzkaller’s built-in classification labels. Only crashes with confirmed kernel-related causes were retained for subsequent statistical counting in our evaluations.  This issue has rarely been addressed explicitly in prior syzkaller-based studies, which renders the crash reports of these studies difficult to directly compare. 

Additionally, certain crash types may exhibit persistent triggering behavior under a fixed experimental environment, with occurrence frequencies significantly exceeding those of others. This skew can distort the distribution of experimental data and bias subsequent crash analysis. To mitigate this effect, we leveraged Syzkaller’s built-in threshold mechanism, capping the number of occurrences of each crash type at 100 within a single experiment. 

{\bf Syscall evolution across kernel versions}. As the kernel evolves, syscall specifications undergo certain changes, such as modifications to parameter types or value ranges, and adjustments to sanitizers of these parameters in syscalls. This can result in legitimate syscall sequences or kernel crashes obtained on one kernel version becoming inapplicable to another version.  In our evaluations, $DTN$ is derived from the DongTing dataset, whose syscall sequences were obtained from kernels older than version 5.17, while our experiments indicated that Psyzkaller can significantly improve fuzzing performance across all considered kernel versions.  Figure~\ref{fig:crash-str} further reveals that the proportion of crashes discovered by $DTN$ on kernel version 5.19 relative to the total crashes found by Psyzkaller is significantly higher than that on other kernel versions.  Thus, learning SDRs embedded in the execution sequences of the current kernel version is crucial for detecting vulnerabilities in the latest kernel. Psyzkaller's approach of simultaneously learning from execution data of current and historical kernels helps to sustain its vulnerability detection capability on future kernel versions.

Additionally, to eliminate potential confounding effects from kernel version discrepancies, all comparative analyses are conducted on the same kernel version(s). That is, in all cases of comparative analysis, a unified baseline is enforced, with consistent kernel versions as a prerequisite.

{\bf The contingency of fuzzing}.  To accurately validate the effectiveness of our approach, fuzzing needs a long run time. Due to the randomness of fuzzing, its results may vary significantly. Therefore, multiple runs are required to average out the random effect. However, there is no consensus on the number of repetitions and the duration of each run required across the academia. Therefore, we chose the norm followed by many prior research: all experiments were run for 48 hours and repeated three times with the average taken. 



\section{Related Work}\label{sec:related-work}
Extensive research has aimed to optimize OS kernel fuzzing in general, and Syzkaller in particular, from various perspectives. Two areas most closely related to our work are \emph{Syscall Specification Generation} and \emph{SDR Extraction}. 

\textbf{Syscall Specification Generation.} Research in this category focuses on developing more accurate and comprehensive syscall specifications to improve seed construction. DIFUZE~\cite{corina_difuze:_2017} uses static analysis of kernel code to automatically infer descriptions of device interfaces for seed generation.  FuzzNG~\cite{bulekov_no_2023} traces kernel execution to construct more accurate syscall specifications. IMF~\cite{han_imf:_2017} employs static analysis to build usage models for IOKitLib syscalls in macOS.  SyzGen~\cite{chen_syzgen_2021} combines dynamic instrumentation and static analysis to generate interface specifications for proprietary device drivers in macOS.  SyzDescribe~\cite{syzdescribe} analyzes kernel modules to acquire their priorities and then examines syscall handlers to recover information such as command values and argument types, thereby enabling more accurate Linux driver specifications.  SyzGen++~\cite{syzgen++} leverages symbolic execution on syscall interfaces to explicitly identify dependencies between syscalls operating on the same kernel objects (e.g., file handlers), improving the accuracy of syscall specifications. KernelGPT~\cite{yang_kernelgpt_2025} synthesizes syscall specifications via Large Language Models (LLMs). These works are orthogonal to ours: they focus on generating precise syscall specifications/templates, which can improve the syntactic compliance of generated seeds and refine Syzkaller’s Choice Table by eliminating spurious SDRs inferred from inaccurate syscall templates.  

\textbf{SDR Extraction.} Work closest to ours focuses on extracting or learning SDRs to improve kernel fuzzing, though only a few efforts exist.  HFL~\cite{kim_hfl_2020} proposes hybrid fuzzing that augments fuzzing with symbolic execution to analyze implicit SDRs and parameter relations.  KernInst~\cite{hao_demystifying_2022} collects SDRs across kernel modules from fuzzing history and refines them using binary analysis, though its efficiency is limited by extensive manual intervention to solve SDR constraints.  Healer~\cite{sun_healer_2021} refines SDRs encoded in Syzkaller’s Choice Table by iteratively removing syscalls from generated sequences and assessing their impact on coverage.  MOCK~\cite{mock} minimizes syscall sequences that trigger new coverage to train a neural language model, which then guides seed mutation in a context-aware manner.  ACTOR~\cite{actor} learns SDRs by constructing \emph{action}s which represent the relations among  syscalls operating on shared objects. Then it generates seeds by assembling actions according to manually-written vulnerability-triggering templates.

Our approach differs from these methods by leveraging Bigram models to learn SDRs from both historical kernel execution data and fuzzing results on the fly. This design offers a balance of \emph{interpretability}, \emph{computational efficiency}, and \emph{fine-grained control} over the relative influence of normal and abnormal executions during SDR learning, thereby enabling Psyzkaller to balance exploration with vulnerability targeting more effectively.

\section{Conclusion}\label{sec:Conclusion}
We have presented an approach of leveraging Bigram models to learn SDRs from both historical and on-the-fly kernel execution data to guide seed generation in Syzkaller for better fuzzing effectiveness.  A key direction for future work is to augment N-gram-based learning with additional sources of information, such as syscall source code and documentation, to extract more accurate and semantically meaningful SDRs.  Beyond seed generation, the learned SDRs could also inform other stages of kernel fuzzing—such as seed mutation, scheduling, and crash clustering—potentially in combination with existing optimization techniques.

\section*{Acknowledgment}
\addcontentsline{toc}{section}{Acknowledgment}
This work is supported in part by National Cryptologic Science Fund of China under Grant No. 2025NCSF02021 and the National Natural Science Foundation of China under Grant No. 62072448.




\bibliographystyle{IEEEtran}
\bibliography{IEEEabrv, SI, kernel, urls}

\end{document}